\documentclass{osa-article}

\journal{osajournal}
\usepackage{multirow}
\usepackage{bm}

\articletype{letter}
\begin{document}
\title{Measurement of the Kerr nonlinear refractive index and its variation among 4H-SiC wafers}

\author{Jingwei Li,\authormark{1}, Ruixuan Wang \authormark{1}, Lutong Cai \authormark{1}, and Qing Li\authormark{*1}}

\address{\authormark{1}Department of Electrical and Computer Engineering, Carnegie Mellon University, Pittsburgh, PA 15213, USA}

\email{\authormark{*}qingli2@andrew.cmu.edu} 


\begin{abstract}
The unique material property of silicon carbide (SiC) and the recent demonstration of low-loss SiC-on-insulator integrated photonics platform have attracted considerable research interests for chip-scale photonic and quantum applications. Here, we carry out a thorough investigation of the Kerr nonlinearity among 4H-SiC wafers from several major wafer manufacturers, and reveal for the first time that their Kerr nonlinear refractive index can be significantly different. By eliminating various measurement errors in the four-wave mixing experiment and improving the theoretical modeling for high-index-contrast waveguides, the best Kerr nonlinear refractive index of 4H-SiC wafers is estimated to be approximately four times of that of stoichiometric silicon nitride in the telecommunication band. In addition, experimental evidence is developed that the Kerr nonlinearity in 4H-SiC wafers can be stronger along the $c$-axis than that in the orthogonal direction, a feature that was never reported before.    
\end{abstract}
\noindent 
\section{Introduction}
Silicon carbide (SiC) recently emerged as a promising photonic and quantum material due to its unique properties, including a wide transparency window spanning from the visible to the mid-infrared, simultaneously possessing second- and third-order optical nonlinearities, large thermal conductivity, and the existence of various color centers that can be exploited as single-photon sources or quantum memories \cite{Awschalom_SiC_qubit, SiC_colorcenter_review, Vuckovic_SiC_review}. In addition, SiC is a robust, CMOS-compatible material with its quality supported by a fast-growing industry, as single-crystal 4H-SiC substrates up to six inches are already commercially available at an affordable cost \cite{Kimoto_SiC_review}. These features, coupled with the recent demonstration of low-loss SiC-on-insulator integrated photonics platform \cite{Lin_3CSiC, Adibi_3CSiC, Noda_4HSiC_PhC, Vuckovic_4HSiC_nphoton,  Ou_4HSiC_combQ}, portend potential disruption of quantum information processing through scalable integration of SiC-based spin defects with a wealth of quantum electrical and photonic technologies on the same chip \cite{Vuckovic_SiC_review}. 

Despite the impressive progresses made in SiC photonics over the past decade, some of its important photonic properties are yet to be fully explored \cite{Ou_SiC_review}. For example, the Kerr nonlinear refractive index $n_2$ of SiC, a third-order nonlinear property that underpins optical nonlinear applications such as optical parametric oscillation (OPO) and Kerr frequency comb generation, is predominantly reported in the literature to be in the range of $(5-8) \times 10^{-19}\ \text{m}^2/\text{W}$ in the telecommunication band (see Table 1). (Note this number is approximately 2-3 times of that of stoichiometric silicon nitride, which is around $2.5\times10^{-19}\ \text{m}^2/\text{W}$ at 1550 nm.) However, our recent work suggested that 4H-SiC wafers from different manufacturers seem to yield different levels of Kerr nonlinearity, as $n_2$ of 4H-SiC from ST Microelectronics (formerly known as Norstel AB and hereinafter referred to as "Norstel" for short) is estimated to be near $(3.0\pm 1.0)\times 10^{-19}\ \text{m}^2/\text{W}$ for the transverse-electric (TE) modes while that of II-VI Incorporated ("II-VI" for short) 4H-SiC wafers is even lower \cite{Li_4HSiC_comb}. A closer look into the literature also exposes the limited data points relied upon by most of the existing works for the $n_2$ estimation, which tended to ignore various uncertainties in the experiment and thus introduced sizeable errors to the process \cite{Lin_3CSiC_nonlinear, 3C_FWM, Gaeta_4HSiC_nonlinear, Ou_4HSiC, Vuckovic_4HSiC_MIcomb, Li_4HSiC_comb}. 

In this work, a systematic approach for the accurate measurement of the Kerr nonlinearity in 4H-SiC wafers is developed. We focus on on-axis, semi-insulating 4H-SiC wafers from three major wafer manufacturers, i.e., Norstel, II-VI, and Cree. While both Cree and Norstel SiC wafers are of high purity (i.e., undoped), the II-VI wafers attain high resistivity through vanadium doping, which has been shown to result in color centers that emit single photons in the telecommunication \textit{O} band (1278-1388 nm) \cite{Awschalom_SiC2}. Our study confirms, for the first time, that the Kerr nonlinearities of the aforementioned commercial 4H-SiC wafers are indeed significantly different, with Cree wafers exhibiting the highest $n_2$ of $(9.1\pm 1.2)\times 10^{-19}\ \text{m}^2/\text{W}$ while II-VI wafers exhibiting the lowest $n_2$ of $(2.3\pm 0.5)\times 10^{-19}\ \text{m}^2/\text{W}$. For 4H-SiC wafers, our work also points to a stronger Kerr nonlinearity along the $c$-axis compared to the orthogonal direction, with the Norstel 4H-SiC wafers exhibiting $n_2$ of $(4.6\pm 0.6)\times 10^{-19}\ \text{m}^2/\text{W}$ for the transverse-magnetic (TM, dominant electric field along the $c$-axis) modes and $n_2$ of $(3.1\pm 0.5)\times 10^{-19}\ \text{m}^2/\text{W}$ for the TE modes (electric field in the wafer plane). Finally, our examination of various waveguide geometries made of the same SiC material also compels an important correction to the existing model for the $n_2$ estimation in high-index-contrast waveguides; otherwise considerable errors can be introduced.  
\begin{table}[ht]
\centering
\begin{tabular}{c c c c c c}
\hline
\textbf{References} & \textbf{SiC} & \textbf{Wafer} & \textbf{Estimation} & $\bm \lambda$ & \textbf{Kerr} $\bm{n_2}$\ \\
& \textbf{polytype} & \textbf{mfr.} & \textbf{method}& \textbf{(nm)} & ($\bm{\times 10^{-19}\ \text{m}^2/\text{W}}$)\\
\hline
 Lu et.al.(2014) \cite{Lin_3CSiC_nonlinear}&a-3C & - & XPM &1550&$5.9\pm0.7$\\ \hline
 Martini et.al. (2018) \cite{3C_FWM} &3C& -&FWM&1550&$5.31\pm0.04$\\
 \hline
 Cardenas et.al.(2015) \cite{Gaeta_4HSiC_nonlinear} &4H&Norstel&SPM&2360&$8.6\pm1.1$ \\
 \hline
 Zheng et.al.(2019) \cite{Ou_4HSiC} &4H&-&FWM&1550&$6.0\pm0.6$\\
 \hline
 Guidry et.al. (2020) \cite{Vuckovic_4HSiC_MIcomb, Vuckovic_4HSiC_soliton} &4H&Cree&OPO&1550&$6.9\pm1.1$\\
 \hline
 Cai et.al. (2022) \cite{Li_4HSiC_comb} &4H&Norstel&Comb&1550&$\perp c: 3.0\pm 1.0$\\
 \hline
 &4H&\textbf{II-VI}&FWM&1550&$\textbf{2.3}\pm \textbf{0.5}$\\
 \cline{2-6}
 \multirow{2}{*}{\textbf{This work}}&\multirow{2}{*}{4H}&\multirow{2}{*}{\textbf{Norstel}}&\multirow{2}{*}{FWM}&\multirow{2}{*}{1550}&$\bm{\perp c}:\textbf{3.1}\pm\textbf{0.5}$\\&&&&&$\bm{// c}: \textbf{4.6}\pm\textbf{0.6}$\\
 \cline{2-6}
&4H&\textbf{Cree}&FWM&1550&$\textbf{9.1}\pm \textbf{1.2}$\\
 \hline
 \end{tabular}
 \caption{Comparison of the measured Kerr nonlinear refractive index of different SiC wafers in the literature versus this work, where various approaches, including cross-phase modulation (XPM), self-phase modulation (SPM), four-wave mixing (FWM), optical parametric oscillation (OPO), and comb generation, were employed. Our work also reveals the larger Kerr nonlinear refractive index for Norstel 4H-SiC wafers along the $c$-axis ($//c$) compared to the orthogonal direction ($\perp c$) for the first time. The Cree and II-VI wafers show similar behavior, although their $n_2$ difference between the two polarizations is smaller and within the measurement uncertainties (see Supplementary).}
\label{Table1}
\end{table}

\section{FWM experiment and $\gamma$ measurement}
Our approach to determining the Kerr nonlinear refractive index is based on measuring the four-wave mixing (FWM) efficiency between two narrow-linewidth lasers (pump and signal, linewidth < 100 kHz) in high-$Q$ SiC microresonators (intrinsic $Q$s in the range of $1-5$ million) \cite{Ho_FWM, 3C_FWM, Ou_4HSiC}. For this purpose, 4-inch-size SiC-on-insulator (SiCOI) wafers were fabricated using a customized bonding and polishing approach (NGK Insulators) for on-axis, semi-insulating 4H-SiC substrates obtained from Norstel, II-VI and Cree (see Supplementary for their wafer specifications). After dicing each wafer to $1\ \text{cm}\times 1\ \text{cm}$ chips, we fabricate high-$Q$ SiC microring and racetrack resonators using ebeam lithography and dry etching. In addition, grating couplers are designed to facilitate the input and output coupling between fibers and on-chip waveguides, with typical insertion loss near 5-7 dB at the center wavelength for each grating coupler \cite{Li_4HSiC_comb}.  
\begin{figure}[ht]
\centering
\includegraphics[width=0.9\linewidth]{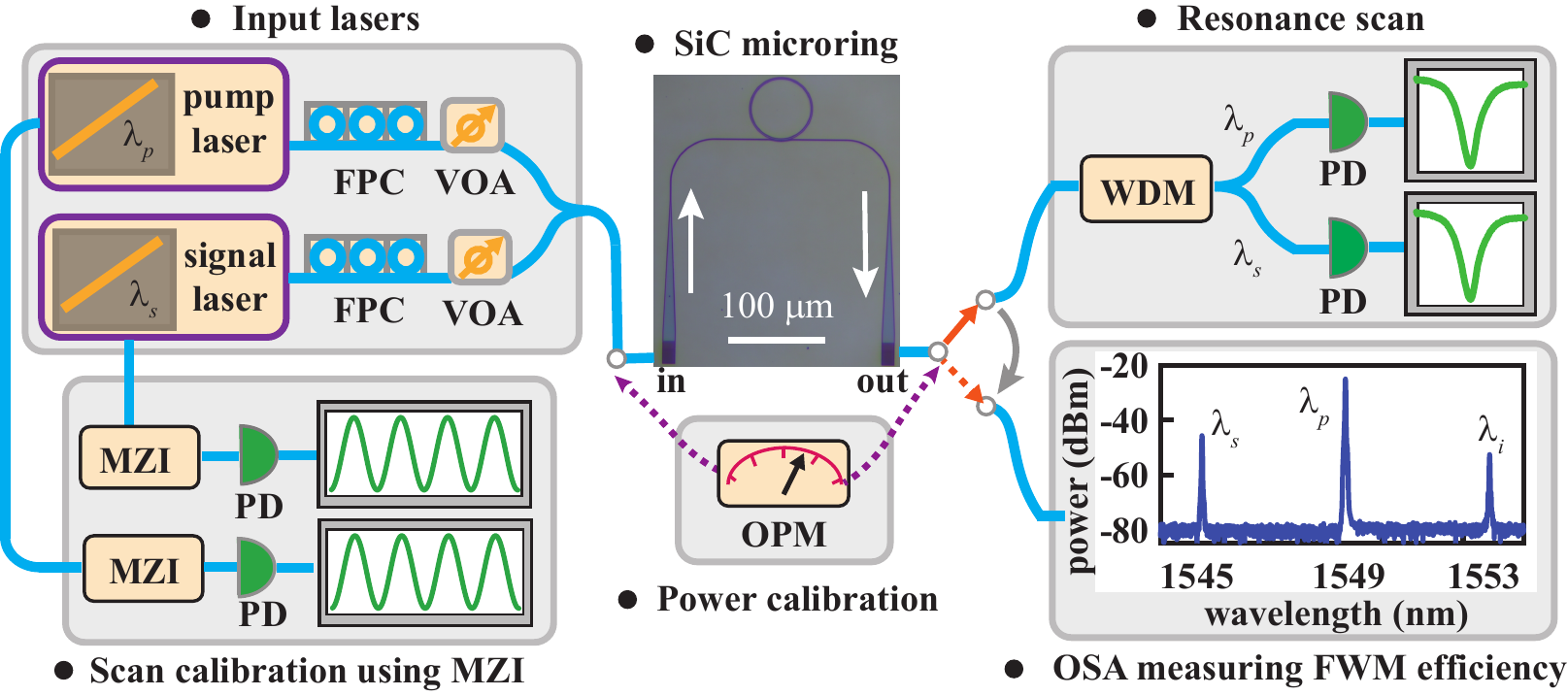}
\caption{Experimental schematic for the measurement of the Kerr nonlinearity in SiC microresonators: FPC, fiber polarization controller; VOA: variable optical attenuator; WDM: wavelength-division multiplexer; PD: photo-detector; MZI: Mach-Zehnder interferometer; OPM: optical power meter; and OSA: optical spectrum analyzer. Detailed description of the experiment is referred to the main text.}
\label{Fig_Schematic}
\end{figure}

As illustrated in Fig.~1, light from the pump laser (Toptica CTL1550, output power fixed at 10 mW) and the signal laser (Agilent 81642A, output power fixed at 1 mW) is combined before being coupled to the on-chip waveguide through a fiber V-groove array (VGA) \cite{Li_4HSiC_comb}. The power of each laser can be externally varied through a variable optical attenuator (VOA) to minimize thermo-optic bistability and higher-order idler generation in the FWM experiment. In addition, the high attenuation accuracy and repeatability (error $<0.1$ dB) of VOAs enables an individual estimation of the on-chip power for the pump and signal separately. This is achieved by applying the maximum attenuation (60 dB) to the pump (signal) laser while keeping the normal attenuation level (<15 dB) for the signal (pump) laser, measuring the off-chip powers from the VGA fibers ("in" and "out" ports as illustrated in Fig.~1) using an optical power meter (OPM), and inferring the corresponding on-chip signal (pump) power with the estimated insertion loss. At the output, the pump and signal wavelengths are separated into two paths through a wavelength-division multiplexing (WDM) filter, allowing each of them to be photo-detected and tuned to their respective resonances from the transmission scan \cite{Li_FWMBS}. Once aligning the pump and signal laser wavelengths to the selected cavity resonances, we measure the idler power, which is generated from the FWM process in the SiC microresonator, using an optical spectrum analyzer (OSA). At this stage, we also tune the pump/signal laser out of resonance and verify that the power measured by OSA is consistent with the number obtained previously from OPM. Such a power calibration scheme proves to be critical as the insertion loss from the chip can deteriorate by 1-2 dB due to unstable fiber-grating alignment during the resonance scan and/or the idler power measurement, resulting in an inaccurate estimation of on-chip powers. 

We define the FWM efficiency as the ratio between the idler power (denoted as $P_i$, which is the on-chip idler power in the waveguide) and the signal power (denoted as $P_{s,\text{in}}$, which is the on-chip signal power before entering the SiC microresonator). In the frequency matched scenario, i.e., the pump, signal and idler are all perfectly aligned to their respective resonances and their wavelengths are close to each other, this FWM efficiency is given by the following expression \cite{Ho_FWM}:
\begin{equation}
\frac{P_i}{P_{s,\text{in}}}=\left(\frac{2\lambda_p}{\pi n_g\sqrt{L}}\right)^4\cdot \left(\gamma P_{p,\text{in}}\right)^2\cdot \left(\frac{Q_l^2}{Q_c} \right)^2_p\left(\frac{Q_l^2}{Q_c} \right)_s\left(\frac{Q_l^2}{Q_c} \right)_i, \label{Eq_1}
\end{equation}
where $\lambda_p$ is the pump wavelength; $n_g$ is the group index of the resonant modes in the C band; $L$ is the circumference of the SiC microresonator; $\gamma$ is the FWM nonlinear parameter which is proportional to the Kerr nonlinear refractive index $n_2$; $P_{p,\text{in}}$ denotes the on-chip pump power before entering the SiC microresonator; and $Q_l$ ($Q_c$) is the loaded (coupling) $Q$ of the resonant mode with the subscripts $p,s,i$ denoting the pump, signal, and idler, respectively. According to Eq.~\ref{Eq_1}, $\gamma$ is explicitly determined by the following factors:
\begin{equation}
\gamma=\left(\frac{\pi n_g \sqrt{L}}{2\lambda_p}\right)^2\cdot \sqrt{\frac{P_i}{P_{s,\text{in}}}}\cdot \frac{1}{P_{p,\text{in}}} \cdot\sqrt{\left(\frac{Q_c}{Q_l^2}\right)_p^2\left(\frac{Q_c}{Q_l^2}\right)_s\left(\frac{Q_c}{Q_l^2}\right)_i}, \label{Eq_2}
\end{equation}
where the first multiplying factor can be accurately computed given that $\lambda_p$ and $L$ are known, and $n_g$ is inferred from the mode's free spectral range (FSR, which is related to $n_g$ through $\text{FSR}=c/(n_gL)$ with $c$ being the speed of light in vacuum). The second multiplying factor in Eq.~\ref{Eq_2}, which is the ratio between the on-chip idler power (after the SiC microresonator) and signal power (before the SiC microresonator), is experimentally determined by tuning the pump laser into resonance and recording the idler power (when the signal is on resonance) and the signal power (when it is off resonance) both from OSA (see Fig.~1). This practice removes uncertainty in the common loss factor shared by the signal and idler, including the insertion loss from the grating coupler and fiber connectors. To address the possibility that this loss factor might be slightly different between the signal and idler, we switch their spectral positions (i.e., set the signal laser at the idler wavelength while keeping the pump the same) and obtain another FWM efficiency for statistical averaging. As such, the FWM efficiency can be reliably measured with an estimated relative uncertainty $<10\%$. The third factor in $\gamma$ is inversely proportional to the on-chip power for the pump, whose error is predominantly caused by the unstable fiber-grating alignment during the FWM experiment. With our power calibration protocol in place (see discussions following Fig.~1), its relative uncertainty is controlled to be $<10\%$. The final constituent factor in $\gamma$ indicates the crucial importance of accurate $Q$ estimation, as $\gamma$ scales as $Q_c^2/Q_l^4$ and a $10\%$ error in $Q_l$ can generate up to $20\%-40\%$ errors in the $\gamma$ estimation. 

\begin{figure}[ht]
\centering
\includegraphics[width=0.98\linewidth]{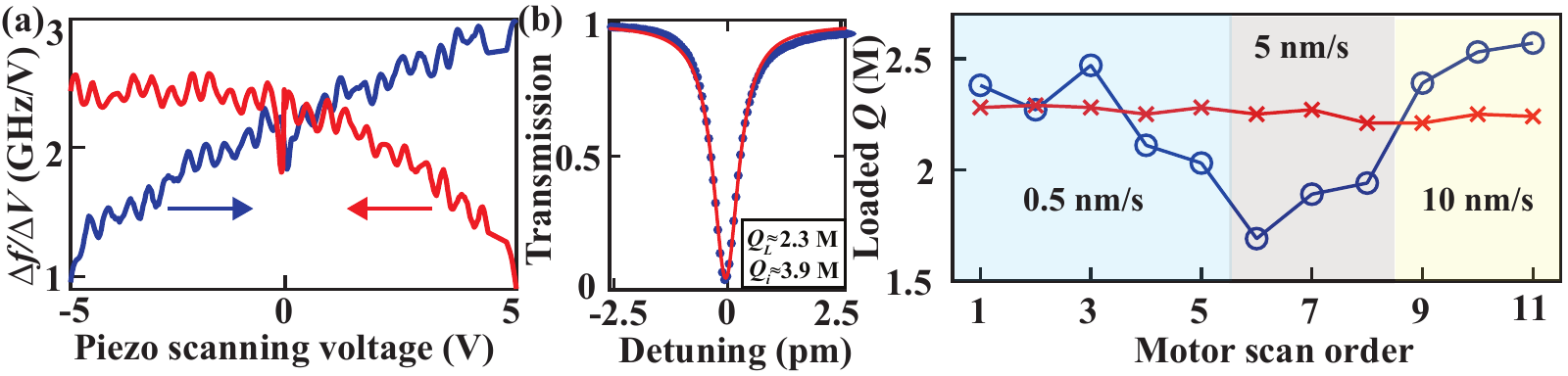}
\caption{(a) Non-uniform frequency tuning rate in the piezo scan of the signal laser (Agilent 81642A) characterized by an imbalanced Mach-Zehnder interferometer (MZI, see Fig.~1): the blue and red curves correspond to the forward and backward scan responses at a scan rate of 1 Hz, respectively. (b) Left: Swept-wavelength transmission of a representative high-$Q$ resonance in a SiC racetrack microresonator with a bending radius of $100\ \mu$m (TM$_{00}$ in Fig.~3(d)): the blue dots are the experimental data and the red curve is its Lorentzian fitting, showing a loaded (intrinsic) $Q$ near 2.3 (3.9) million; Right: Extracted loaded $Q$s for the same resonance shown on the left with repeated continuous sweeps from the signal laser. The three colored regions indicate the different tuning speeds varied from 0.5 nm/s to 10 nm/s with all the other scanning parameters kept the same: the blue circles are the loaded $Q$s extracted directly from motor scans (fluctuations up to $20\%$) and the red crosses are the $Q$s calibrated using MZI (fluctuations $<3\%$).}
\label{Fig_Qscan}
\end{figure}

To accurately determine the $Q$ factors from the linear swept-wavelength transmission measurement, we divide a portion of the tunable laser output to a fiber-based MZI, which has a path difference of three meters and an FSR of 68.1 MHz around 1550 nm (see Fig.~1). By scanning the SiC chip and MZI simultaneously and using the known FSR of the MZI to calibrate the swept wavelengths, we are able to correct various scan nonidealities arising from the limited tuning resolution in tunable lasers. Take the signal laser (Agilent 81642A) for example: the frequency tuning rate of the piezo scan (i.e., varying the laser frequency in a narrow range by applying an external voltage) is found to be nonuniform across a linear voltage scan (Fig.~2(a)). This directly affects the $Q$ estimation as the inferred cavity linewidth will depend on the relative position of the resonance within the scan range, which is difficult to control precisely from one scan to another. On the other hand, repeated continuous frequency sweeps from the laser's motor scan also yield $10\%-20\%$ fluctuations in the inferred loaded $Q$s without calibration (Fig.~2(b)). Such scan nonidealities are ultimately related to the limited wavelength resolution (pm level) present in most of tunable lasers, which poses a challenge to determining optical $Q$s accurately on the million level and above. Hence, the introduction of the MZI to this experiment for the scan calibration becomes necessary, which improves the uncertainty in the $Q_l$ estimation to be $<3\%$ (Fig.~2(b)). 

\begin{figure}[ht]
\centering
\includegraphics[width=0.95\linewidth]{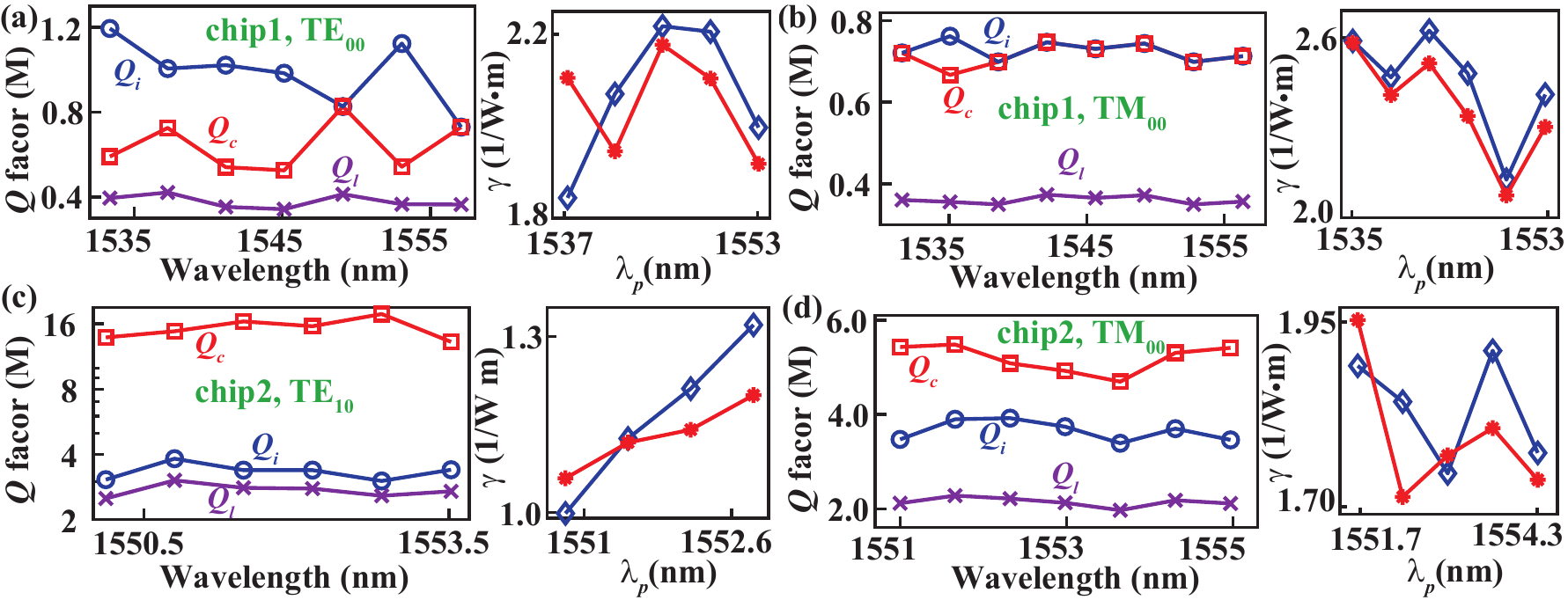}
\caption{Experimental results for the $\gamma$ estimation in four devices from two distinct Norstel SiC chips. For (a)-(d): the left figure shows the measured loaded $Q$ ($Q_l$) as well as inferred intrinsic $Q$ ($Q_i$) and coupling $Q$ ($Q_c$) for various azimuthal orders of the resonances used in the FWM experiment (note the $y$ axis for (c) is in the log scale while the rest is linear). On the right we plot the extracted $\gamma$ for the corresponding pump wavelengths, with the signal and idler resonances being 1 FSR away. The blue diamond and red star curves are for the same pump resonance but with the signal and idler positions exchanged. Devices in (a) and (b) (from chip 1) are 36-$\mu$m-radius microrings while devices for (c) and (d) (from chip 2) are racetrack resonators with a bending radius of 100 $\mu$m. Their specific waveguide geometries are provided in Table \ref{Table2}.}
\label{Fig_Gamma}
\end{figure}

Despite the developed calibration processes for the power and $Q$ measurement, appreciable variations (on the order of $20\%-30\%$) in the $\gamma$ estimation (and hence $n_2$) still exist. To further reduce the uncertainties, we carry out the FWM experiment on multiple devices for each SiC material so that a statistically meaningful average is obtained. Moreover, different combinations of azimuthal orders in each device are employed to account for the variations in their intrinsic and coupling $Q$s, which are partially attributed to their scattering-limited radiation losses and frequency-dependent couplings \cite{Li_azimuth}. In Fig.~3, exemplary results for four different devices based on the Norstel SiC are presented: the two devices corresponding to 
 Figs.~3(a) and 3(b) are 36-$\mu$m-radius microrings from the SiC chip that has been previously used for the microcomb generation \cite{Li_4HSiC_comb}, with an approximate SiC thickness around 475 nm; on the other hand, the devices corresponding to Figs.~3(c) and 3(d) are larger racetrack resonators (bending radius of $100\ \mu$m and circumference near 1.3 mm),  which are fabricated on a different SiC chip with a nominal thickness around 850 nm. To ensure frequency matching between the interacting waves in the FWM process, we choose resonances belonging to the same mode family with only one FSR separation and verify that their dispersion is indeed small enough \cite{Li_4HSiC_comb}. The mode order and polarization of each mode family are identified by comparing the measured FSR and coupling $Q$s to the simulation results \cite{Li_azimuth}. While in theory we should expect a uniform $\gamma$ for the same mode family, the fluctuations observed in Fig.~3 indicate that the aforementioned experimental uncertainties for the $\gamma$ estimation cannot be completely removed.  

\section{$n_2$ estimation from measured $\gamma$}
 After extracting $\gamma$ from the FWM experiment for each device, the final step in the Kerr nonlinear refractive index measurement is to connect $\gamma$ to $n_2$ based on $\gamma=2\pi n_2/(\lambda_p A_\text{eff})$, where $A_\text{eff}$ is the effective mode area. The exact definition of $A_\text{eff}$, however, is not well agreed upon in the literature. For example, one common version of $A_\text{eff}$ that is applicable to low-index-contrast waveguides takes the following form \cite{Agrawal_NLF}:
 \begin{equation}
 A_\text{eff}=\frac{\left (\iint_{-\infty}^{\infty} |\bm{E}(x,y)|^2 dxdy\right )^2}{\iint_{\text{core}} |\bm E(x,y)|^4 dxdy},
 \label{Eq_A1}
 \end{equation}
where $\bm E(x,y)$ is the electric field of the waveguide mode under consideration and $x, y$ are the coordinates in the waveguide cross-section. (Note the denominator in Eq.~\ref{Eq_A1} is only integrated within the waveguide core, which is the only material assumed to possess a nonzero $n_2$.) For high-index-contrast waveguides, which is the case for SiCOI, we believe that $A_\text{eff}$ needs to be modified as (see derivation in Sect.~III.D of the Supplementary from Ref.~\cite{Li_FWMBS}):
\begin{equation}
\tilde{A}_{\text{eff}}= \frac{\left(\iint_{-\infty}^{\infty} \epsilon_r(x,y) |\bm E(x,y)|^2\ dxdy \right)^2}{\iint_{\text{core}}\epsilon_r^2(x,y) |\bm E(x,y)|^4\ dxdy}\cdot \left(\frac{n_0}{n_g}\right )^2,
\label{Eq_A2}
\end{equation}
where $\epsilon_r(x,y)$ is the relative permittivity and $n_0$ denotes the refractive index of the waveguide core. Note that while the first multiplying factor in Eq.~\ref{Eq_A2} resembles the effective mode volume derived in Ref.~\cite{Painter_Kerr}, an additional correcting factor, which depends on the ratio between $n_0$ and $n_g$ (group index), is introduced here. This factor can be intuitively understood based on the fact that $n_2$ is defined for the bulk material while $\gamma$ is obtained from confined waveguide modes. 

Aside from theoretical justification, experimental evidence for the correct $A_\text{eff}$ can be developed by computing $n_2$ from the measured $\gamma$ for various waveguide geometries made of the same material, which should result in a consistent $n_2$. Such an example is provided in Table 2 for the SiC devices measured in Fig.~3. By focusing on the TM polarization, we find that Eq.~\ref{Eq_A1} resulted in dramatically different numerical values of $n_2$ for the two distinct waveguide geometries corresponding to Figs.~3(b) and 3(d), despite the fact that they are both fabricated from the same Norstel SiC wafer. In contrast, the application of Eq.~\ref{Eq_A2} leads to consistent $n_2$ (within measurement uncertainties) for a variety of waveguide geometries (more evidence in Supplementary), which lends strong support to its validity. A closer look into the two $A_\text{eff}$ formulas suggests (see Supplementary) that Eq.~\ref{Eq_A1} is only applicable when the waveguide mode is well confined within the core and the corresponding group index $n_g$ is similar to the refractive index of the bulk material $n_0$ (e.g., the TE modes in Fig.~3); otherwise the more generic formula, i.e, Eq.~\ref{Eq_A2}, should be used for the $n_2$ estimation. 

\begin{table}
\centering
\begin{tabular}{c c c c c c c c }
\hline
\textbf{Norstel}&\multirow{2}{*}{\textbf{Mode}}& \textbf{Width} & \textbf{Height}&\textbf{Measured} $\bm \gamma$ & $\bm{n_2}$ \textbf{with Eq.~\ref{Eq_A1}}&$\bm{n_2}$ \textbf{with Eq.~\ref{Eq_A2}}\\
devices& &(nm)& (nm) & $1/(\text{W}\cdot \text{m})$&($\times 10^{-19}\ \text{m}^2/\text{W}$)&($\times 10^{-19}\ \text{m}^2/\text{W}$)\\
\hline
Fig.~3(a)&TE$_{00}$&$2200\pm100$&$475\pm25$&$2.05\pm0.15$&$3.9\pm 0.6$ &$3.1\pm0.5$\\
\hline
Fig.~3(b)&TM$_{00}$&$2500\pm100$&$475\pm25$&$2.4\pm0.2$& $10.0\pm2.0$&$4.6\pm0.6$\\
\hline
Fig.~3(c)&TE$_{10}$&$2500\pm 100$&$850\pm50$&$1.15\pm 0.1$ & $4.0\pm0.6$&$3.5\pm 0.6$\\
\hline
Fig.~3(d)&TM$_{00}$&$2500\pm100$&$850\pm50$&$1.8\pm 0.1$&$6.3\pm 0.8 $&$5.3\pm0.8$\\

\hline
\end{tabular}
\caption{Estimation of the Kerr nonlinear refractive index and the impact of different $A_\text{eff}$ formulas for SiC devices shown in Fig.~3, all of which were made from the same Norstel SiC material.}
\label{Table2}
\end{table}

Given the sensitivities of the $\gamma$ estimation to the $Q$ measurement and the smaller uncertainties in the $Q$ estimation of 36-$\mu$m-radius microrings compared to those of the larger racetrack resonators, we adopt the $n_2$ result for the Norstel material in Table 1 based on Figs.~3(a) and 3(b). In the Supplementary, we provide additional experimental data for the Cree and II-VI SiC wafers and summarize their results in Table 1, both of which are based on the FWM measurement in 36-$\mu$m-radius microrings. We want to emphasize that one of the main conclusions of this work, that the Kerr nonlinear refractive index $n_2$ from the three major SiC wafer manufacturers is significantly different, is unlikely to be caused by the errors introduced in the connection from the experimentally measured $\gamma$ to $n_2$. This is because we can focus on the TE-polarized modes that are well confined in the in-plane direction (waveguide widths $>2\ \mu$m), for which different $A_\text{eff}$ expressions yield similar results (see Supplementary for a table summary for the TE modes). 
\section{Conclusion}
In conclusion, we developed a systematic approach for the accurate measurement of the Kerr nonlinearity in 4H-SiC wafers, and showed, for the first time, that there are significant variations in the Kerr nonlinear refractive index among 4H-SiC wafers from different manufacturers. Our work also revealed a larger Kerr nonlinearity along the $c$-axis than that in the orthogonal direction, and a necessary correction in the modeling of $n_2$ to obtain consistent results in high-index-contrast waveguides. We believe these findings, in particular the fact that the Kerr nonlinear refractive index of 4H-SiC can be up to four  times that of stoichiometric silicon nitride, are crucial to the future development of the SiCOI platform for a variety of nonlinear applications in both the classical and quantum regimes.

\newpage
\title{Supplementary Material}
\setcounter{equation}{0}
\setcounter{figure}{0}
\setcounter{table}{0}
\setcounter{section}{0}
\renewcommand{\thetable}{S\arabic{table}}
\renewcommand{\thefigure}{S\arabic{figure}}
\renewcommand{\theequation}{S\arabic{equation}}
\noindent 
\section{4H-SiC wafer specification}
We list the wafer specifications of 4-inch-size, semi-insulating 4H-SiC wafers that have been used in this work in the following table:

\begin{table}
\centering
\begin{tabular}{c c c c c c c c }
\hline
\textbf{Wafer}&\multirow{2}{*}{\textbf{Grade}}& \textbf{Orientation} & \textbf{MPD}&\textbf{TTV} &\textbf{Bow} & \textbf{Warp} & \textbf{Resistivity}\\
\textbf{mfr.}& &(deg)&$(\text{cm}^{-2}$)&($\mu$m)&$(\mu$m)&($\mu$m)&($\Omega \cdot \text{cm})$\\
\hline
II-VI&Prime&0.02&0.1&1.3&4.6&0.5&$3\times10^{11}$\\
\hline
Norstel&Test&0.05&0.5&1.8&-5&17&$1.7\times10^{9}$\\
\hline
Cree&Production&-&-&-&-&-&$>5\times10^{5}$\\
\hline
\end{tabular}
\caption{Wafer specifications of 4-inch-size, semi-insulating 4H-SiC wafers obtained from the three major SiC wafer manufacturers. MPD: micropipe density. }
\label{TableS_Material}
\end{table}

It is worth noting that for II-VI 4H-SiC wafers, optical tests from multiple ($>5$) wafers and in different batches confirm that their optical properties, including the Kerr nonlinearity, are fairly consistent with no noticeable changes. For the Norstel SiC, we only managed to obtain two wafers of the test grade. Due to their significant warp values, uneven SiC thicknesses (up to 100 nm variations) are observed in the device layer following the bonding and polishing step \cite{Li_4HSiC_comb}. The Cree data is based on SiC chips made from a single Cree wafer of the production grade (which does not seem to have an inspection report).     

\section{$\gamma$ measurement and $n_2$ estimation for Cree devices}
We perform similar device fabrication and four-wave mixing (FWM) measurements for the Cree SiC wafer as we did in the main text for the Norstel material. The Cree chip has an estimated thickness of ($630 \pm 30$) nm based on reflectometry. The SiC microrings have a radius of $36\ \mu$m and varied ring widths. In the dry etching step, we remove approximately 500 nm SiC (calibrated using profilometer), leaving a pedestal layer with a nominal thickness around 130 nm. In the end of the fabrication, a 1-$\mu$m-thick PECVD oxide layer is deposited on top of the SiC devices.  
\begin{figure}[ht]
\centering
\includegraphics[width=0.95\linewidth]{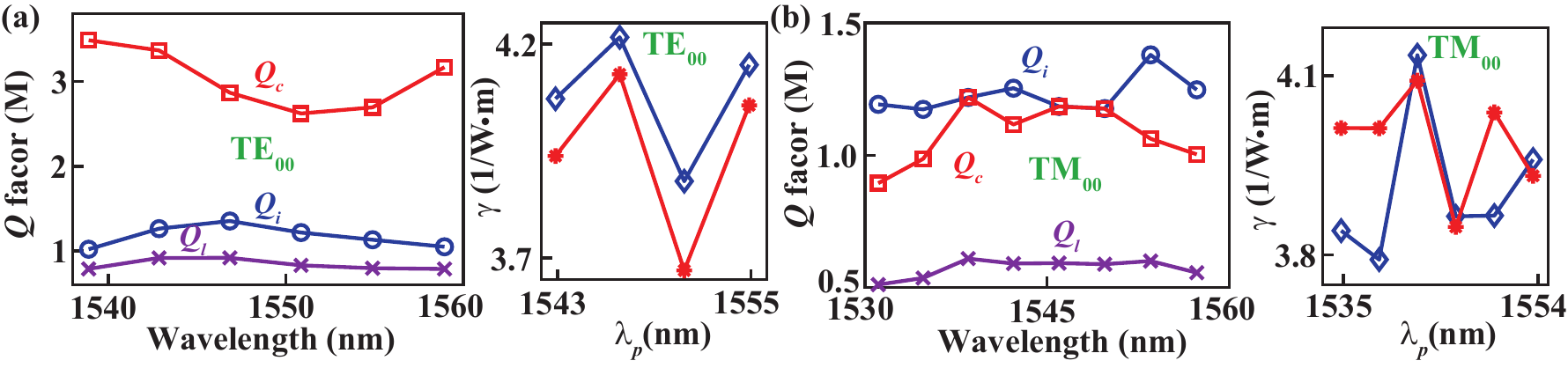}
\caption{Experimental results for the $\gamma$ estimation in two 36-$\mu$m-radius SiC microrings on a Cree SiC chip. The specific device parameters are listed in Table \ref{TableS_Cree}. For (a) and (b): the left figure shows the measured loaded $Q$ ($Q_l$) as well as inferred coupling $Q$ ($Q_c$) and intrinsic $Q$ ($Q_i$) for the resonances that have been used in the FWM experiment (pump, signal and idler are only separated by 1 FSR); and on the right we plot the extracted $\gamma$ for varied pump wavelengths (i.e., different azimuthal orders), with the blue diamond (red star) curve corresponding to the case that the signal wavelength is smaller (larger) than the pump wavelength.}
\label{FigS_Cree}
\end{figure}

\begin{table}[h]
\centering
\begin{tabular}{c c c c c c c c }
\hline
\textbf{Cree}&\multirow{2}{*}{\textbf{Mode}}& \textbf{Width} & \textbf{Height}&\textbf{Measured} $\bm \gamma$ & $\bm{n_2}$ \textbf{with Eq.~3}&$\bm{n_2}$ \textbf{with Eq.~4}\\
devices& &(nm)& (nm) & $1/(\text{W}\cdot \text{m})$&($\times 10^{-19}\ \text{m}^2/\text{W}$)&($\times 10^{-19}\ \text{m}^2/\text{W}$)\\
\hline
Fig.~\ref{FigS_Cree}(a)&TE$_{00}$&$2500\pm100$&$630\pm30$&$4.0\pm0.18$&$10.4\pm 1.2$ &$8.9\pm1.1$\\
\hline
Fig.~\ref{FigS_Cree}(b)&TM$_{00}$&$2500\pm100$&$630\pm30$&$3.95\pm0.11$& $13.6\pm1.2$&$9.4\pm0.9$\\
\hline
\end{tabular}

\caption{Estimation of the Kerr nonlinear refractive index for the Cree SiC devices shown in Fig.~\ref{FigS_Cree}. Both devices have an etch depth near 500 nm and a top cladding layer of oxide. The sidewall angle of the device is estimated to be near 80 degrees.}
\label{TableS_Cree}
\end{table}
In Fig.~\ref{FigS_Cree}, we present exemplary results for the TE and TM resonances supported by the SiC microrings. Their polarization and mode order are identified by comparing the measured FSRs and coupling $Q$s to the simulation results \cite{Li_azimuth}. Using the extracted $\gamma$, we estimate $n_2$ in Table \ref{TableS_Cree} by taking the uncertainties in the waveguide dimensions into consideration. While the mean value of $n_2$ for the TM polarization (whose dominant electric field is along the $c$-axis) is slightly bigger than that of the TE polarization (whose dominant electric field is orthogonal to the $c$-axis), this difference ($\approx 5\%$) is within the measurement error and is not statistically significant. Therefore, we averaged $n_2$ for the TE and TM polarizations in Table 1 of the main text and increased its uncertainty slightly to account for both cases.

\section{$\gamma$ measurement and $n_2$ estimation for II-VI devices}
Likewise, we fabricate 36-$\mu$m-radius SiC microrings on semi-insulating II-VI 4H-SiC (primary grade) chips and perform FWM experiments to extract their $\gamma$ and $n_2$. The II-VI chip has an estimated SiC thickness of ($600 \pm 30$) nm based on reflectometry. In the dry etching process, we remove approximately 500 nm SiC, leaving a pedestal layer with a nominal thickness around 100 nm. For this chip, the top cladding is air.  
\begin{figure}[ht]
\centering
\includegraphics[width=0.95\linewidth]{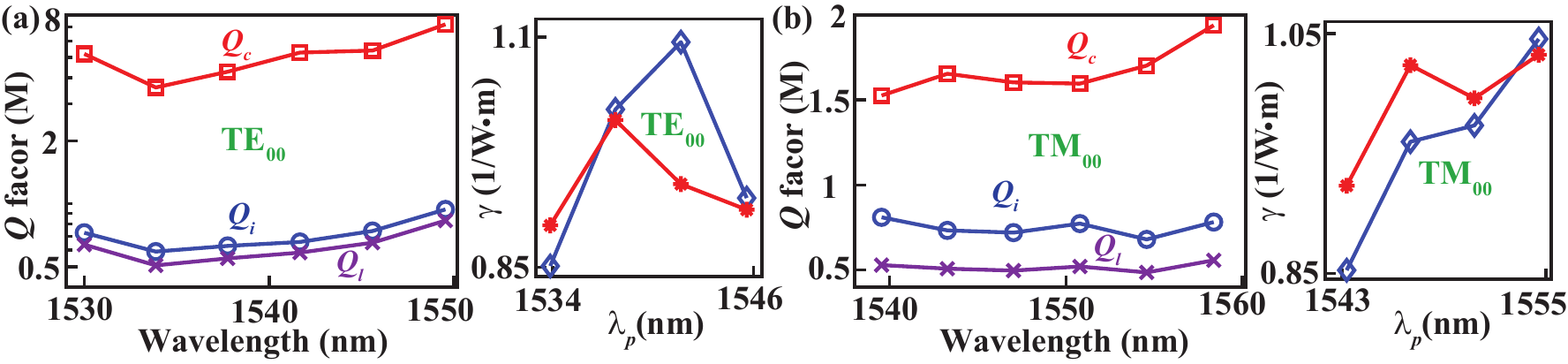}
\caption{Experimental results for the $\gamma$ estimation in two devices from the same II-VI SiC chip. The specific device parameters are listed in Table \ref{TableS_IIVI}. For (a) and (b): the left figure shows the measured loaded $Q$ ($Q_l$) as well as inferred coupling $Q$ ($Q_c$) and intrinsic $Q$ ($Q_i$) for the resonances that have been used in the FWM experiment (pump, signal and idler are only separated by 1 FSR); and on the right we plot the extracted $\gamma$ for varied pump wavelengths (i.e., different azimuthal orders), with the blue diamond (red star) curve corresponding to the case that the signal wavelength is smaller (larger) than the pump wavelength. Note that the $y$ axis in (a) is in the log scale as the coupling $Q$s of the TE$_{00}$ mode family are much larger than the intrinsic $Q$s (i.e., under-coupled).}
\label{FigS_IIVI}
\end{figure}

\begin{table}[h]
\centering
\begin{tabular}{c c c c c c c c }
\hline
\textbf{II-VI}&\multirow{2}{*}{\textbf{Mode}}& \textbf{Width} & \textbf{Height}&\textbf{Measured} $\bm \gamma$ & $\bm{n_2}$ \textbf{with Eq.~3}&$\bm{n_2}$ \textbf{with Eq.~4}\\
devices& &(nm)& (nm) & $1/(\text{W}\cdot \text{m})$&($\times 10^{-19}\ \text{m}^2/\text{W}$)&($\times 10^{-19}\ \text{m}^2/\text{W}$)\\
\hline
Fig.~\ref{FigS_IIVI}(a)&TE$_{00}$&$2500\pm100$&$600\pm30$&$0.96\pm0.08$&$2.3\pm 0.3$ &$2.0\pm0.3$\\
\hline
Fig.~\ref{FigS_IIVI}(b)&TM$_{00}$&$3000\pm100$&$600\pm30$&$0.98\pm0.07$& $3.9\pm0.4$&$2.5\pm0.4$\\
\hline
\end{tabular}
\caption{Estimation of the Kerr nonlinear refractive index for the II-VI SiC devices shown in Fig.~\ref{FigS_IIVI}. Note both devices have an etch depth near 500 nm and a top cladding of air. The sidewall angle of the device is estimated to be near 80 degrees.}
\label{TableS_IIVI}
\end{table}

In Fig.~\ref{FigS_IIVI}, we present representative results for the TE and TM resonances supported by the SiC microrings. As can be seen, the mean value of $n_2$ along the $c-$axis (TM) is approximately $20\%-30\%$ larger than that of the orthogonal direction (TE). Nevertheless, this difference is still within the measurement uncertainties. As such, we took the averaged $n_2$ for the TE and TM polarizations in Table 1 of the main text, and increased its uncertainty to account for both cases.

\section{Detailed comparison of different $A_\text{eff}$ expressions}
In this section, we will take a closer look into the two $A_\text{eff}$ formulas that were discussed in the main text. For convenience, we reproduce their expressions below:
\begin{equation}
 A_\text{eff}=\frac{\left (\iint_{-\infty}^{\infty} |\bm{E}(x,y)|^2 dxdy\right )^2}{\iint_{\text{core}} |\bm E(x,y)|^4 dxdy},
 \label{EqS_A1}
 \end{equation}
 and 
 \begin{equation}
\tilde{A}_{\text{eff}}= \frac{\left(\iint_{-\infty}^{\infty} \epsilon_r(x,y) |\bm E(x,y)|^2\ dxdy \right)^2}{\iint_{\text{core}}\epsilon_r^2(x,y) |\bm E(x,y)|^4\ dxdy}\cdot \left(\frac{n_0}{n_g}\right )^2,
\label{EqS_A2}
\end{equation}
where $\bm E(x,y)$ is the electric field of the waveguide mode under consideration; $x, y$ are the coordinates in the waveguide cross-section; $\epsilon_r(x,y)$ is the relative permittivity; and $n_0$ denotes the refractive index of the waveguide core. 

In the literature, Eq.~\ref{EqS_A1} (Eq.~3 in the main text) is commonly used for the $\gamma$ calculation from $n_2$ as $\gamma=2\pi n_2/(\lambda_p A_\text{eff})$ \cite{Agrawal_NLF}. As explained in the main text, we believe that a modified formula, i.e., Eq.~\ref{EqS_A2} (Eq.~4 in the main text), is required for high-index-contrast waveguides \cite{Li_FWMBS}. By comparing the inferred $n_2$ from these two expressions for varied waveguide geometries made of the same material, we find that Eq.~\ref{EqS_A2} provides a consistent estimation of $n_2$, which lends strong support to its validity. By contrast, results based on Eq.~\ref{EqS_A1} often lead to dramatic variations in $n_2$ that are difficult, if not impossible, to explain for high-quality, single-crystal materials used in this work. 

\begin{figure}[ht]
\centering
\includegraphics[width=0.95\linewidth]{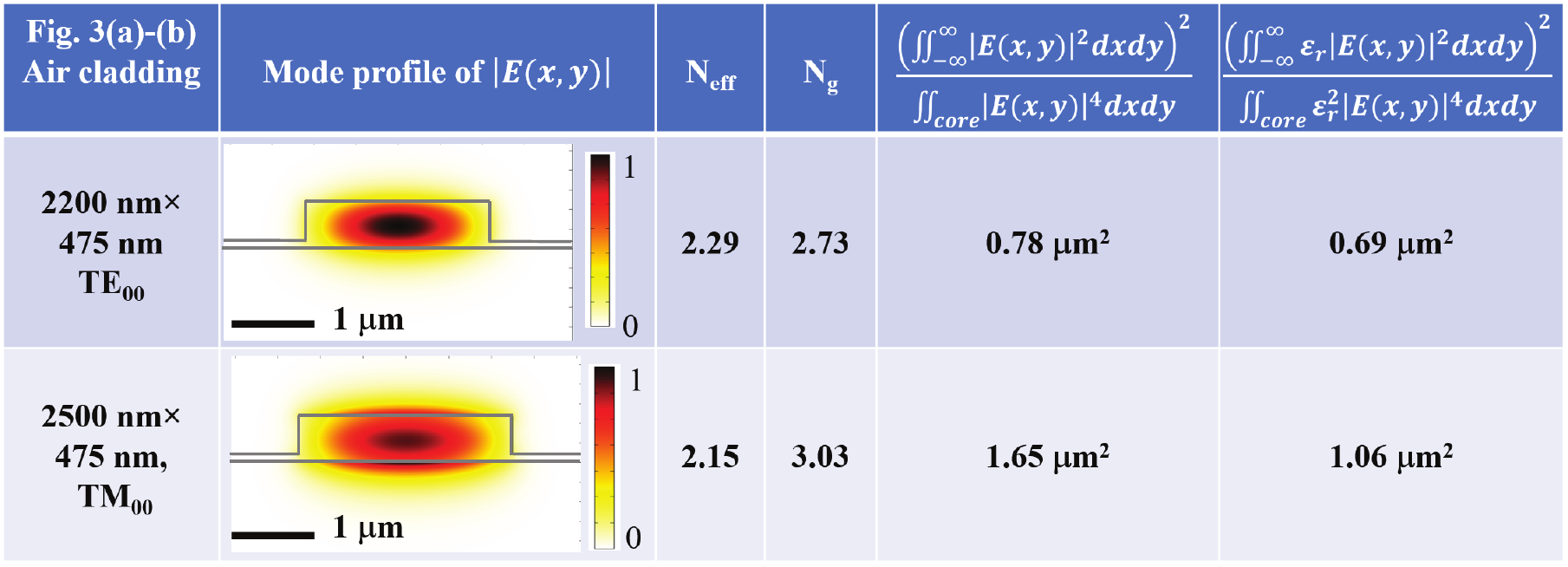}
\caption{Computation of two different expressions of $A_\text{eff}$, i.e., Eq.~\ref{EqS_A1} and Eq.~\ref{EqS_A2}, for the waveguide modes corresponding to Figs.~3(a) and 3(b) in the main text. While numerical values of $A_\text{eff}$ for the TE polarization are reasonably close between the two formulas, their results are more than two times different for the TM polarization, which are contributed by the weighted field integral and a factor depending on the ratio between $n_g$ and $n_0$. Both waveguides have an oxide cladding underneath and an air cladding on top.}
\label{FigS_Aeff}
\end{figure}

To better understand the difference between the two $A_\text{eff}$ expressions, in particular their reasonable agreement for the TE-polarized modes and significant disagreement for the TM-polarized modes in Table 2 (of the main text), we use the waveguides modes corresponding to Figs.~3(a) and 3(b) (of the main text) as an example. As shown in Fig.~\ref{FigS_Aeff}, the TE modes are well confined within the waveguide core and their group index is close to the material index $n_0$ ($n_0\approx 2.56$ at 1550 nm). As a result, the difference between Eqs.~\ref{EqS_A1} and \ref{EqS_A2} is relatively small. On the other hand, the TM mode expands more outside the waveguide core, given that the vertical dimension is much smaller than the horizontal dimension. This results in a $35\%$ reduction in the field integral of $A_\text{eff}$ by weighting the electric field with the relative permittivity (i.e., $\epsilon_r$), as done in Eq.~\ref{EqS_A2}, compared to the one without (as in Eq.~\ref{EqS_A1}). In addition, Eq.~\ref{EqS_A2} has another multiplying factor that depends on the ratio between $n_0$ and $n_g$. Because the group index $n_g$ for the TM mode is considerably larger than $n_0$, this factor will contribute another $30\%$ reduction in the effective mode area. Combined together, the numerical value of $A_\text{eff}$ given by Eq.~\ref{EqS_A2} is approximately $46\%$ of that obtained with Eq.~\ref{EqS_A1} for the waveguide mode corresponding to Fig.~3(b). We believe the data presented in this paper unanimously supports the adoption of Eq.~\ref{EqS_A2} (Eq.~4 in the main text) as the general formula for connecting $\gamma$ to $n_2$, while Eq.~\ref{EqS_A1} (Eq.~3 in the main text) is only applicable for waveguide mode that is well confined in the waveguide core and whose group index is similar to the refractive index of the bulk material.  

\section{Summary table for $n_2$ comparison based on TE modes}
\begin{table}[h]
\centering
\begin{tabular}{c c c c c c c c }
\hline
\textbf{SiC}&\multirow{2}{*}{\textbf{Fig.~}}& \textbf{Width} & \textbf{Height}&\textbf{Measured} $\bm \gamma$ & $\bm{n_2}$ \textbf{with Eq.~3}&$\bm{n_2}$ \textbf{with Eq.~4}\\
\textbf{mfr.}& &(nm)& (nm) & $1/(\text{W}\cdot \text{m})$&($\times 10^{-19}\ \text{m}^2/\text{W}$)&($\times 10^{-19}\ \text{m}^2/\text{W}$)\\
\hline
II-VI&S2(a)&$2500\pm100$&$600\pm30$&$0.96\pm0.08$&$2.3\pm 0.3$ &$2.0\pm0.3$\\
\hline
Norstel&3(a)&$2200\pm100$&$475\pm25$&$2.05\pm0.15$&$3.9\pm 0.6$ &$3.1\pm0.5$\\
\hline
Cree&S1(a)&$2500\pm100$&$630\pm30$&$4.0\pm0.18$&$10.4\pm 1.2$ &$8.9\pm1.1$\\
\hline
\end{tabular}
\caption{Summary of the experimental results for the TE$_{00}$ mode family in 36-$\mu$m-radius SiC microrings made from semi-insulating, on-axis 4H-SiC wafers from three major wafer manufacturers. The two $A_\text{eff}$ expressions (i.e., Eq.~3 and Eq.~4 in the main text) provide a reasonably close estimation of $n_2$ for each material, confirming that its numerical values are indeed significantly different among 4H-SiC wafers produced by II-VI, Norstel and Cree.}
\label{TableS_compareTE}
\end{table}

\noindent Finally, we want to emphasize that one of the main conclusions of this work, that the Kerr nonlinear refractive index $n_2$ from the three major SiC wafer manufacturers is significantly different, is unlikely to be caused by the errors introduced in the connection from the experimentally measured $\gamma$ to $n_2$. This is because we can focus on the TE-polarized modes that are well confined in the in-plane direction (waveguide widths $>2\ \mu$m), for which different $A_\text{eff}$ expressions yield similar results. For this purpose, we summarize the experimental results corresponding to the TE$_{00}$ mode family in 36-$\mu$m-radius SiC microrings and the estimated $n_2$ in Table \ref{TableS_compareTE}.

\begin{backmatter}
\bmsection{Funding}
This work was supported by DARPA (D19AP00033) and NSF (2127499). 

\bmsection{Acknowledgments}
The authors would like to thank the helpful discussions with Prof.~Robert Davis from CMU and equipment support from Dr.~Lijun Ma and Dr.~Oliver Slattery at NIST. J. Li also acknowledges the support of Axel Berny Graduate Fellowship from CMU. 

\bmsection{Disclosures}  The authors declare no conflicts of interest.

\bmsection{Data Availability} A subset of data underlying the results presented in this paper is provided in the Supplementary. More comprehensive data can be obtained from the authors upon request.

\end{backmatter}


\bibliography{SiC_Ref}

\begin{thebibliography}{10}
\newcommand{\enquote}[1]{``#1''}

\bibitem{Awschalom_SiC_qubit}
C.~P. Anderson, A.~Bourassa, K.~C. Miao, G.~Wolfowicz, P.~J. Mintun, A.~L.
  Crook, H.~Abe, J.~U. Hassan, N.~T. Son, T.~Ohshima, and D.~D. Awschalom,
  \enquote{Electrical and optical control of single spins integrated in
  scalable semiconductor devices,} {\protect\JournalTitle{Science}}
  \textbf{366}, 1225--1230 (2019).

\bibitem{SiC_colorcenter_review}
S.~Castelletto and A.~Boretti, \enquote{Silicon carbide color centers for
  quantum applications,} {\protect\JournalTitle{Journal of Physics: Photonics}}
  \textbf{2}, 022001 (2020).

\bibitem{Vuckovic_SiC_review}
D.~M. Lukin, M.~A. Guidry, and J.~Vu{\v c}kovi{\'c}, \enquote{Integrated
  {{quantum photonics}} with {{silicon carbide}}: challenges and
  {{prospects}},} {\protect\JournalTitle{PRX Quantum}} \textbf{1}, 020102
  (2020).

\bibitem{Kimoto_SiC_review}
T.~Kimoto, \enquote{Material science and device physics in {SiC} technology for
  high-voltage power devices,} {\protect\JournalTitle{Japanese Journal of
  Applied Physics}} \textbf{54} (2015).

\bibitem{Lin_3CSiC}
X.~Lu, J.~Y. Lee, P.~X.-L. Feng, and Q.~Lin, \enquote{Silicon carbide microdisk
  resonator,} {\protect\JournalTitle{Optics Letters}} \textbf{38}, 1304--1306
  (2013).

\bibitem{Adibi_3CSiC}
T.~Fan, H.~Moradinejad, X.~Wu, A.~A. Eftekhar, and A.~Adibi,
  \enquote{High-{{Q}} integrated photonic microresonators on
  {{3C}}-{{SiC}}-on-insulator ({{SiCOI}}) platform,}
  {\protect\JournalTitle{Optics Express}} \textbf{26}, 25814--25826 (2018).

\bibitem{Noda_4HSiC_PhC}
B.-S. Song, T.~Asano, S.~Jeon, H.~Kim, C.~Chen, D.~D. Kang, and S.~Noda,
  \enquote{Ultrahigh-{{Q}} photonic crystal nanocavities based on {{4H}}
  silicon carbide,} {\protect\JournalTitle{Optica}} \textbf{6}, 991 (2019).

\bibitem{Vuckovic_4HSiC_nphoton}
D.~M. Lukin, C.~Dory, M.~A. Guidry, K.~Y. Yang, S.~D. Mishra, R.~Trivedi,
  M.~Radulaski, S.~Sun, D.~Vercruysse, G.~H. Ahn, and J.~Vu{\v c}kovi{\'c},
  \enquote{{{4H}}-silicon-carbide-on-insulator for integrated quantum and
  nonlinear photonics,} {\protect\JournalTitle{Nature Photonics}} \textbf{14},
  330--334 (2020).

\bibitem{Ou_4HSiC_combQ}
C.~Wang, Z.~Fang, A.~Yi, B.~Yang, Z.~Wang, L.~Zhou, C.~Shen, Y.~Zhu, Y.~Zhou,
  R.~Bao, Z.~Li, Y.~Chen, K.~Huang, J.~Zhang, Y.~Cheng, and X.~Ou,
  \enquote{High-{{Q}} microresonators on {{4H}}-silicon-carbide-on-insulator
  platform for nonlinear photonics,} {\protect\JournalTitle{Light: Science \&
  Applications}} \textbf{10}, 139 (2021).

\bibitem{Ou_SiC_review}
A.~Yi, C.~Wang, L.~Zhou, Y.~Zhu, S.~Zhang, T.~You, J.~Zhang, and X.~Ou,
  \enquote{Silicon carbide for integrated photonics,}
  {\protect\JournalTitle{Applied Physics Reviews}} \textbf{9}, 031302 (2022).
  Publisher: American Institute of Physics.

\bibitem{Li_4HSiC_comb}
L.~Cai, J.~Li, R.~Wang, and Q.~Li, \enquote{Octave-spanning microcomb
  generation in {4H}-silicon-carbide-on-insulator photonics platform,}
  {\protect\JournalTitle{Photonics Research}} \textbf{10}, 870--876 (2022).

\bibitem{Lin_3CSiC_nonlinear}
X.~Lu, J.~Y. Lee, S.~Rogers, and Q.~Lin, \enquote{Optical {{Kerr}} nonlinearity
  in a high-{{Q}} silicon carbide microresonator,}
  {\protect\JournalTitle{Optics Express}} \textbf{22}, 30826 (2014).

\bibitem{3C_FWM}
F.~Martini and A.~Politi, \enquote{Four wave mixing in {3C} {SiC} ring
  resonators,} {\protect\JournalTitle{Applied Physics Letters}} \textbf{112},
  251110 (2018).

\bibitem{Gaeta_4HSiC_nonlinear}
J.~Cardenas, M.~Yu, Y.~Okawachi, C.~B. Poitras, R.~K.~W. Lau, A.~Dutt, A.~L.
  Gaeta, and M.~Lipson, \enquote{Optical nonlinearities in high-confinement
  silicon carbide waveguides,} {\protect\JournalTitle{Optics Letters}}
  \textbf{40}, 4138--4141 (2015).

\bibitem{Ou_4HSiC}
Y.~Zheng, M.~Pu, A.~Yi, X.~Ou, and H.~Ou, \enquote{{{4H}}-{{SiC}} microring
  resonators for nonlinear integrated photonics,} {\protect\JournalTitle{Optics
  Letters}} \textbf{44}, 5784 (2019).

\bibitem{Vuckovic_4HSiC_MIcomb}
M.~A. Guidry, K.~Y. Yang, D.~M. Lukin, A.~Markosyan, J.~Yang, M.~M. Fejer, and
  J.~Vu{\v c}kovi{\'c}, \enquote{Optical parametric oscillation in silicon
  carbide nanophotonics,} {\protect\JournalTitle{Optica}} \textbf{7}, 1139
  (2020).

\bibitem{Awschalom_SiC2}
G.~Wolfowicz, C.~P. Anderson, B.~Diler, O.~G. Poluektov, F.~J. Heremans, and
  D.~D. Awschalom, \enquote{Vanadium spin qubits as telecom quantum emitters in
  silicon carbide,} {\protect\JournalTitle{Science Advances}} \textbf{6},
  eaaz1192 (2020).

\bibitem{Vuckovic_4HSiC_soliton}
M.~A. Guidry, D.~M. Lukin, K.~Y. Yang, R.~Trivedi, and J.~Vučković,
  \enquote{Quantum optics of soliton microcombs,} {\protect\JournalTitle{Nature
  Photonics}} \textbf{16}, 52--58 (2022).

\bibitem{Ho_FWM}
P.~P. Absil, J.~V. Hryniewicz, B.~E. Little, P.~S. Cho, R.~A. Wilson, L.~G.
  Joneckis, and P.-T. Ho, \enquote{Wavelength conversion in {GaAs} micro-ring
  resonators,} {\protect\JournalTitle{Optics Letters}} \textbf{25}, 554--556
  (2000).

\bibitem{Li_FWMBS}
Q.~Li, M.~Davan{\c c}o, and K.~Srinivasan, \enquote{Efficient and low-noise
  single-photon-level frequency conversion interfaces using silicon
  nanophotonics,} {\protect\JournalTitle{Nature Photonics}} \textbf{10},
  406--414 (2016).

\bibitem{Li_azimuth}
Q.~Li, A.~A. Eftekhar, Z.~Xia, and A.~Adibi, \enquote{Azimuthal-order
  variations of surface-roughness-induced mode splitting and scattering loss in
  high-{Q} microdisk resonators,} {\protect\JournalTitle{Optics Letters}}
  \textbf{37}, 1586--1588 (2012).

\bibitem{Agrawal_NLF}
G.~P. Agrawal, \emph{Nonlinear Fiber Optics (Sixth Edition)} (Academic Press,
  2019).

\bibitem{Painter_Kerr}
Q.~Lin, T.~J. Johnson, R.~Perahia, C.~P. Michael, and O.~J. Painter, \enquote{A
  proposal for highly tunable optical parametric oscillation in silicon
  micro-resonators,} {\protect\JournalTitle{Optics Express}} \textbf{16},
  10596--10610 (2008).

\end{thebibliography}

\end{document}